\documentclass[runningheads]{llncs}

\usepackage[ruled,linesnumbered,lined,boxed,commentsnumbered]{algorithm2e}
\usepackage{amsmath}
\usepackage{amssymb}
\usepackage{array}
\usepackage{floatrow} 
\usepackage{graphicx}
\usepackage{multirow}
\usepackage[lofdepth,lotdepth]{subfig}
\usepackage{url}

\newfloatcommand{capbtabbox}{table} [][\FBwidth]

\DeclareMathOperator*{\argmax}{arg\,max}

\urldef{\mail}\path|liuwenqiangcs@gmail.com, liukeen@mail.xjtu.edu.cn, duanhaimeng@gmail.com, whu@nju.edu.cn, weibifan@mail.xjtu.edu.cn|

\begin{document}

\pagestyle{empty}
\mainmatter

\title{Exploiting Source-Object Networks to Resolve Object Conflicts in Linked Data}
\titlerunning{Exploiting Source-Object Network to Resolve Object Conflicts}

\author{Wenqiang Liu$^{1}$ \and Jun Liu$^1$ \and Haimeng Duan$^1$ \and Wei Hu$^2$ \and Bifan Wei$^1$}
\authorrunning{W. Liu et al.}

\institute{$^1$MOEKLINNS Lab, Xi'an Jiaotong University, China\\
$^2$State Key Laboratory for Novel Software Technology, Nanjing University, China\\
\mail}
\maketitle


\begin{abstract} 
Considerable effort has been exerted to increase the scale of Linked Data. However, an inevitable problem arises when dealing with data integration from multiple sources. Various sources often provide conflicting objects for a certain predicate of the same real-world entity, thereby causing the so-called \emph{object conflict} problem. At present, object conflict problem has not received sufficient attention in the Linked Data community. Thus, in this paper, we firstly formalize the object conflict resolution as computing the joint distribution of variables on a heterogeneous information network called the \emph{Source-Object Network}, which successfully captures three correlations from objects and Linked Data sources. Then, we introduce a novel approach based on network effects called \emph{ObResolution} (object resolution), to identify a true object from multiple conflicting objects. ObResolution adopts a pairwise Markov Random Field (pMRF) to model all evidence under a unified framework. Extensive experimental results on six real-world datasets show that our method achieves higher accuracy than existing approaches and it is robust and consistent in various domains.
\end{abstract}


\section{Introduction}
\label{sect:intro}

Considerable effort has been made to increase the scale of Linked Data. Especially, the number of available Linked Data sources in the Linking Open Data (LOD) project increases from 12 in 2007 to 1,146 in 2017.\footnote{\url{http://lod-cloud.net/}} In this paper, a Linked Data source refers to a dataset that has been published to the LOD project by individuals or organizations, such as YAGO. Linked Data resources are encoded in the form of $\langle\textit{Subject},\textit{Predicate}, \textit{Object}\rangle$ triples through the Resource Description Framework (RDF) format. The subject denotes the resource, and predicate is used to express a relationship between subject and object. Inevitably, errors occur during such creation process given that many Linked Data sources on the web have been created from semi-structured datasets (e.g., Wikipedia) and unstructured ones through automatic or semi-automatic algorithms \cite{3}. As a result, a predicate for the same real-world entity can have multiple inconsistent objects when dealing with data integration from multiple sources. For example, the objects of the \textit{dbp:populationTotal} for \textit{Beijing} in Freebase\footnote{\url{https://www.freebase.com/m/072p8}} and DBpedia\footnote{\url{http://dbpedia.org/resource/Beijing}} are ``20,180,000" and ``21,516,000" respectively. In this paper, this problem is called the \emph{object conflict} problem. The concept of object conflicts can be defined as two objects are being in conflict only when their similarity is less than the defined threshold. According to this definition, it is also likely to regard two objects expressed in terms of different measure units as conflicts. But, the purpose of our study is to rank the trust values of all objects and provide the most common ones for users, rather than remove some objects directly. Therefore, people who use our methods can still see all objects. 

A straightforward method to resolve object conflicts is to conduct the majority voting, which regards the object with the highest number of occurrences as the correct object. The drawback of this method is that it assumes that all Linked Data sources are equally reliable \cite{liu2016truthdiscover}. In reality, some Linked Data sources are more reliable than others and thus may produce inaccurate results in scenarios when there are some Linked Data sources provide untrustworthy objects. Many truth discovery methods have been proposed to estimate source reliability \cite{6,liu2016truthdiscover,11,9} in recent years to overcome the limitation of majority voting. The basic principle of these methods is that a source which provides trustworthy objects more often is more reliable, and an object from a reliable source is more trustworthy. Therefore, the truth discovery problem in these methods is formulated as an iterative procedure, which starts by assigning the same trustworthiness to all Linked Data sources, and iterates by computing the trust value of each object and propagating back to the Linked Data sources. 

However, a major problem occurs in the aforementioned approaches. The iterative procedure in these methods is performed by simple weighted voting, which can result in that the rich getting richer over iterations \cite{31}. Especially in Linked Data, data sharing between different Linked Data sources is common in practice. Therefore, errors can easily propagate and lead to wrong objects often appearring in many sources. As a result, methods based on an iterative procedure may derive a wrong conclusion. The situation is even worse for many predicates that are time sensitive, which the corresponding object tends to change over time (e.g., \textit{dbo:populationTotal}), because many out-of-date objects often exist in more Linked Data sources than those up-to-date objects. The experimental results of \cite{liu2016truthdiscover} based on an iterative procedure also show the same conclusion, which obtained the lowest accuracy for the time-sensitive predicate

To address this problem, we propose a new method, called \emph{ObResolution} (Object Resolution), which utilizes the Source-Object network to infer the true object. This network successfully captures three correlations from objects and Linked Data sources. For example, an object from a reliable source is more trustworthy and a source that provides trustworthy objects more often is more reliable. Thus, we build a message propagation-based method that exploits the network structure to infer the trust values of all objects and then the object with the maximum trust score is regarded as the true object. According to our evaluation, our method outperforms several existing truth discovery methods because these methods either model all clues by the iterative procedure, or do not take the sharing between Linked Data sources into consideration. We summarize the main contributions of our work as follows. 
\begin{itemize}
\item We formalize the object conflict resolution problem as computing the joint distribution of variables in a heterogeneous information network called the \emph{Source-Object Network}, which successfully captures three correlations from objects and Linked Data sources.
\item We propose a novel truth discovery approach, ObResolution, to identify the truth in Linked Data. This approach leverages pairwise Markov Random Field (pMRF) to model the interdependencies from objects and sources, and a message propagation-based method is utilized that exploits the Source-Object Network structure to infer the trust values of all objects. 
\item We conducted extensive experiments on six real-world Linked Data datasets to validate the effectiveness of our approach. Our experimental results showed that our method achieves higher accuracy than several comparable methods.
\end{itemize}
The remainder of this paper is organized as follows. Section~\ref{sect:pre} formalizes our problem and the details of our method are discussed in Section~\ref{sect:obr}. The evaluation of our method is reported in Section~\ref{sect:eval}. Related work is discussed in Section~\ref{sect:work}. Section~\ref{sect:concl} presents the conclusion and future work. 

\section{Preliminaries}
\label{sect:pre}

\subsection{Basic Definitions}

\textbf{Definition 1 (RDF Triple)} \cite{30}. We let $I$ denote the set of IRIs (Internationalized Resource Identifiers), $B $ denote the set of blank nodes, and $L $ denote the set of literals (denoted by quoted strings, e.g., ``\textit{Beijing City}"). An RDF triple can be represented by $ \langle s,p,o \rangle  \in (I\cup B)\times I \times (I \cup B \cup L)$, where $s$ is called \textit{subject}, $p$ is \textit{predicate}, and $o$ is \textit{object}. \\

\noindent\textbf{Definition 2 (Trustworthiness of Sources)}. The trustworthiness $t(\omega_j) $ of a source $\omega_j $ is the average probability of the object provided by $\omega_j $ being true as defined as follows:
\begin{equation}
t(\omega_j)=\sum_{o_i \in F(\omega_j)}\tau(o_i)/ |F(\omega_j)|,
\end{equation}
\noindent where $F(\omega_j) $ is the set of objects provided by source $\omega_j $ and $\tau(o_i) $ denotes the trust value of an object $o_i $. \\

\noindent\textbf{Definition 3 (Trust Values of Objects)} \cite{11}. The trust value $\tau(o_i) $ of an object $o_i $ is the probability of being correct,  which can be computed as 
\begin{equation}\label{eq10}
\tau(o_i)=\sum_{\omega_j \in \Omega (o_i)}t(\omega_j)/|\Omega (o_i)|,
\end{equation}
\noindent where $\Omega(o_i)$ represents the set of sources that provide object $o_{i}$.\\

We let $O=\{o_{i}\}_{m} $ denote a set of conflicting objects for a certain predicate of a real-world entity. The process of object conflict resolution in Linked Data is formally defined as follows. Given a set of conflicting objects $O$, \textit{ObResolution} assigns a trust score that lies in between 0 and 1 to each object. A score of object close to $1$ indicates that we are very confident that this object is true. Therefore, the truth can be represented by $o^*=\argmax\limits_{o_{i}\in{O}}\;\tau(o_{i})$.

\subsection{Problem Analysis}
Through the observation and analysis of the object conflicts in our sample Linked Data, we found three helpful correlations from Linked Data sources and objects to effectively distinguish between true and false objects.

\begin{itemize}
\item\textbf{Correlations among Linked Data Sources and Objects.} If an object comes from a reliable source, it will be assigned a high trust value. Thus a source that provides trustworthy objects often has big chance to be selected as a reliable source. For example, the object provided by DBpedia is more reliable than objects supported by many small sources because DBpedia is created from Wikipedia. This condition also serves as a basic principle for many truth discovery methods \cite{8,12,14,6,15,9}.
\item\textbf{Correlations among Objects.} If two objects are similar, they should have similar trust values, which indicates that similar objects appear to have mutually support. For example, we assume that one source claims that the \textit{dbp:height} of \textit{Statue of Liberty} is ``46.0248" and another says that it is ``47". If one of these sources has a high trust value, the other should have a high trust value as well. Meanwhile, if two objects are mutually excluded, they cannot be both true. If one of them has a high trust value, the other should have a low trust value. For instance, if two different sources claim that the \textit{dbp:height} of \textit{Statue of Liberty} are ``93" and ``46.0248" respectively. If the true object is ``46.0248", then ``93" should be a wrong object.
\item\textbf{Correlations among Linked Data Sources.} In many truth discovery methods, the trustworthiness of a source is formulated as the probability of the objects provided by this source being the truth. Therefore, the more same objects two different sources provide, the more similar is the trustworthiness of the two sources. Consider an extreme case when two sources provide the same objects for each predicate, and the trustworthiness of these two sources is the same.
\end{itemize}

As discussed, these three principles can be used to infer the trust values of objects. A key problem for object conflicts resolution is how to model these principles under a unified framework.

\section{ObResolution Method}
\label{sect:obr}

In this section, we formally introduce our proposed method called \textit{ObResolution}, for discovering the most reliable objects from the set of conflicting objects. We first formulate the object conflict resolution problem as the Source-Object network analysis problem, which successfully captures all the correlations from objects and Linked Data sources. Subsequently, a message propagation-based method that exploits the Source-Object network structure is introduced to solve this problem. Finally several important issues that make this method practical are discussed.

\subsection{Model Details}
In general, the input to our problem includes three parts: (i) objects, which are the values of a certain predicate for the same real-world entity; (ii) Linked Data sources, which provide these objects, e.g., Freebase; and (iii) mappings between objects and Linked Data sources, e.g., which Linked Data sources provide which objects for the certain predicate of the same real-world entity. Thus, a set of objects and sources can be structured into a bipartite network. In this bipartite network, source nodes are connected to the object nodes, in which links represent the ``provider" relationships. For ease of illustration, we present example network of six sources and four conflicting objects as shown in Figure 1(a). According to the first principle, an object from a reliable source is more trustworthy, and thus a source that provides trustworthy objects than other sources. The ``provided" relationship between a source and an object also indicates the interdependent relationship between the trust value of the object and the trustworthiness of the source. Besides the ``provider" relationship between the source and object, among objects and among Linked Data sources also have correlations. For instance, because sources $\omega_1, \omega_3, \omega_5$ provide the same object $o_1$ in Figure 1(a), they have a correlation for any two of these three sources. Therefore, the bipartite network in Figure 1(a) can be converted to a heterogeneous information network called the Source-Object Network as shown in Figure 1(b). 
\begin{figure}[!htb]
\centering
\includegraphics[width=1\textwidth]{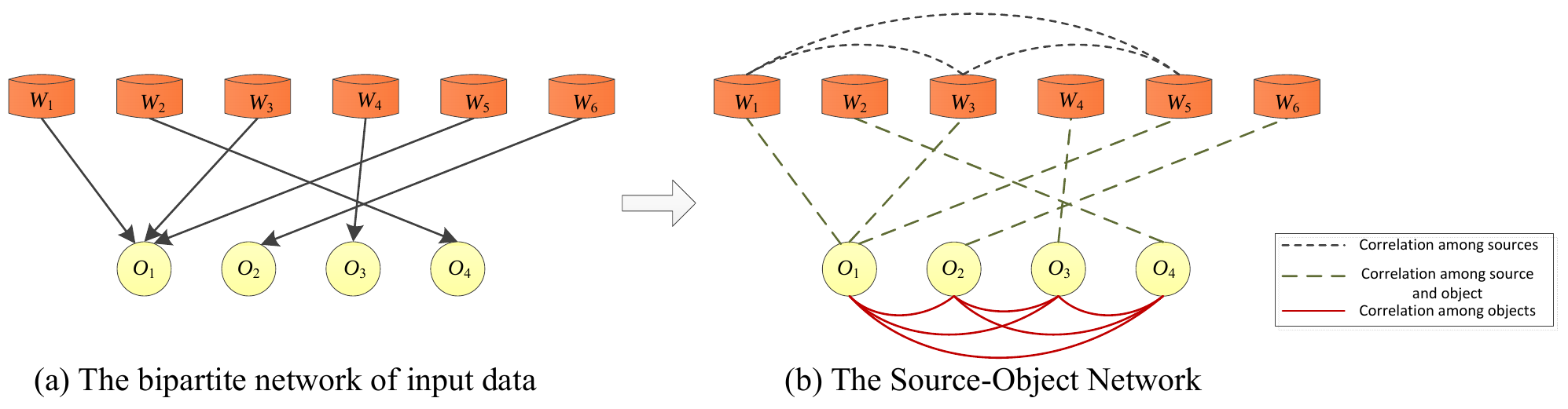}
\caption{Illustration of an example source-Object Network}
\end{figure} 

The Source-Object Network $G=(V,E)$ contains $n$ Linked Data source nodes $\Omega=\{\omega_1,...,\omega_n\}$ and $m$ conflicting object nodes $O=\{o_{1},...,o_{m}\}$, $V=\Omega \cup O$, connected with edge set $E$. Owing to three types of correlations of objects and Linked Data sources, the Source-Object Network $G$ has three types of edges $E=E_{\Omega} \cup E_{O} \cup E_{\Omega\rightarrow O}$, where $E_{\Omega}\subseteq \Omega \times \Omega$ represents the correlations between sources, $E_{O}\subseteq O \times O$ indicates the correlations among objects and $E_{\Omega\rightarrow O}$ represents the ``provided" relationships between sources and objects.

Given a Source-Object Network, which successfully captures three correlations from objects and Linked Data sources, the task is to estimate the reliability of sources and the trust values of all conflicting objects. Each node in $G$ is a random variable that can represent the trust values of objects and trustworthiness of sources. However, we find that the trust values of objects and trustworthiness of sources are assumed to be dependent on their neighbors and independent of all the other nodes in this network. This condition motivates us to select a method based on pMRF, which is a powerful formalism used to model real-world events based on the Markov chain and knowledge of soft constraints. Therefore, the Source-Object Network is represented by pMRF in this study. In fact, pMRF is mainly composed of three components: an unobserved field of random variables, an observable set of random variables, and the neighborhoods between each pair of variables. We let all the nodes $V=\Omega \cup O$ in $G$ be observation variables. Thus, the unobserved variables $Y=Y_\Omega \cup Y_O$ have two types of labels. 

1) The unobserved variable $y_i$ is the label of an object node. It indicates whether the corresponding object is the truth, which follows the Bernoulli distribution defined as follows.
\begin{equation}\label{eq11}
P(y_i) = 
\left\{\begin{matrix}
\tau(o_i) & if \ o_i \ is \ true, i.e., \ y_i=1 , \\ 
 1-\tau(o_i) & \ if \ o_i \ is \ false, i.e., \ y_i=0. 
\end{matrix}\right.
\end{equation}

2) The unobserved variable $y_i$ is the label of 啊 Linked Data source node which represents whether the corresponding source is a reliable source and also follows the Bernoulli distribution.
\begin{equation}\label{eq11}
P(y_j) = 
\left\{\begin{matrix}
t(\omega_j) & if \  \omega_j \ is \ a \ reliable \ source, i.e., \ y_j=1 , \\ 
1-t(\omega_j) & \ \ \ if \ \omega_j \ is \ a \ unreliable \ source, i.e., \ y_j=0. 
\end{matrix}\right.
\end{equation}

The problem of inferring the trust values of conflicting objects and trustworthiness of sources can be converted to compute the joint distribution of variables in pMRF, which is factorized as follows:
\begin{equation}\label{eq12} 
 P(y_1,...,y_m,...,y_{m+n})=\frac{\prod_{c \in \rm C} \psi_c(x_c)}{\sum_{x_c \in X} \prod_{c \in \rm C} \psi_c(x_c)}, 
\end{equation}
where $\rm C$ denotes the set of all maximal cliques, the set of variables of a maximal clique is represented by $x_c$ $(c\in \rm C)$, and $\psi_c(x_c)$ is a potential function in pMRF.

\subsection{Inference Algorithms}
In general, exactly inferencing the joint distribution of variables in pMRF is known to be a non-deterministic polynomial-time hard problem \cite{rayana2015collective}. Loopy Belief Propagation (LBP) is an approximate inference algorithm that has been shown to perform extremely well for various of applications in the real word. In belief propagation, estimating the joint distribution of variables is a process of minimizing the graph energy. The key steps in the propagation process can be concisely expressed below.
\begin{itemize}
\item\textbf{Spreading the Belief Message.} The message from variable $y_i$ to $y_j$ is represented by $m_{i\rightarrow j}(y_j)$, $y_j\in \{0,1\}$, which is defined as follows:
\begin{equation}\label{eq13}
\begin{aligned}
m_{i\rightarrow j}(y_j)\text{=}\sum_{y_i\in \{0,1\}} U(y_i,y_j)\psi_i(y_i)
\prod_{y_k\in N(y_i) \cap  Y\setminus \{y_j\} }m_{k\rightarrow i}(y_i),
\end{aligned}
\end{equation}
\noindent where $N(v_i)$ indicates the set of neighbors of node $y_i $; $\psi_i(y_i)$  denotes the prior belief of $P(y_i)$, and $U(y_i,y_j)$ is a unary energy function. \\

\item\textbf{Belief Assignment.} The marginal probability  $P(y_i)$ of unobserved variable $y_i$ is updated according to its neighbors, and is defined as follows:

\begin{equation}\label{eq14}
P(y_i) =\psi_i(y_i) \prod_{y_j\in N(y_i) \cap  Y}m_{j\rightarrow i}(y_i).
\end{equation}
\end{itemize}

The algorithm updates all messages in parallel and assigns the label until the messages stabilizes, i.e. achieve convergence. Although convergence is not theoretically guaranteed, the LBP has been shown to converge to beliefs within a small threshold fairly quickly with accurate results \cite{rayana2015collective}. After they stabilize, we compute the marginal probability $P(y_i)$. Thus, we can obtain the trust values of object and the trustworthiness of source. Given only one truth for a certain predicate of a real-world entity, the true object is $o_i$ when $\tau(o_i) $ is the maximum. To date, we have described the main steps of LBP, but two problems occur in the algorithm, energy function and prior belief. These problems are discussed as follows.

\textbf{Energy Function.} The energy function $U(y_i,y_j)$ denotes the likelihood of a node with label $y_i$ to be connected to a node with label $y_j$ through an edge. The following three types of energy functions exist depending on the types of edges:
\begin{itemize}
\item The energy function between sources and objects. A basic principle between sources and objects is that the reliable source tends to provide true objects and unreliable sources to false objects. However, a reliable sources may also provide false objects as unreliable sources to true objects. In this study, we let $\beta$ denote the likelihood between reliable sources and true objects, whereas $\delta$ denotes the likelihood between unreliable sources and false objects. Therefore, the energy function between sources and objects is shown in the first three columns of Table \ref{tab:tb1}.
\item The energy function among objects. The more similar the two objects are, the greater is the probability of them having the same trust values. Therefore, a positive correlation exists between the energy function and the similarity $S(o_i,o_j)$ between object $o_i$ and $o_j$, as shown in Table 2.  
\begin{table}[htp]
\begin{floatrow}
\capbtabbox{  
 \begin{tabular}{p{1.5cm}p{1cm}p{1cm}p{1.2cm}p{1.2cm}<{\centering}}
 \hline 
 \multirow{2}{*}{Source} & \multicolumn {2}{c}{Object}& \multicolumn {2}{c}{Source} \\  
 \cline{2-5}
 \centering
 & True&False& Reliable&Unreliable\\  
 \hline
 Reliable& \ \ \,$\beta$ & $1-\beta$ &\ \ \,$\varepsilon$& $1-\varepsilon$\\  
  Unreliable& $1-\delta$& $\ \ \, \delta$ & $1-\varepsilon$&$\varepsilon$\\ 
 \hline  
 \end{tabular}  
}{  
 \caption{Energy function from sources and objects}  
 \label{tab:tb1}  
}  
\capbtabbox{  
  \begin{tabular}{p{1.5cm}p{1.7cm}p{1.7cm}<{\centering}}
 \hline 
 \multirow{2}{*}{Object} & \multicolumn {2}{c}{Object} \\  
 \cline{2-3}
 
 & True&False\\  
 \hline
 True&$\ \ S(o_i,o_j)$  & $1-S(o_i,o_j)$  \\  
  False& $1-S(o_i,o_j)$ & $S(o_i,o_j)$ \\ 
 \hline  
 \end{tabular}   
}{  
\caption{Energy function between objects}  
 \label{tab:tb2}  
}  
\end{floatrow}  
\end{table}  
\vskip -15pt
\item The energy function among sources. We assume that the more same objects two different sources provide, the more similar the trustworthiness of the two sources are. The coefficient $\varepsilon=|F(\omega_i) \cap F(\omega_j)| /max(|F(\omega_i)|, |F(\omega_j)|)$ is used to denote the likelihood between sources $\omega_i $ and $\omega_j$, where $F(\omega_i) $ is the set of objects provided by source $\omega_i $ as shown in the last two columns of Table 1. 
\end{itemize}
The pseudo code of this method is shown in Algorithm~\ref{algo}.
\vskip -15pt
\begin{algorithm}[!htp]
\SetKwData{Left}{left}\SetKwData{This}{this}\SetKwData{Up}{up}
\SetKwFunction{Union}{Union}\SetKwFunction{FindCompress}{FindCompress}
\SetKwInOut{Input}{Input}\SetKwInOut{Output}{Output}
\Input{a set of conflicting objects $O=\{o_{1},...,o_{m}\}$, a set of Linked Data sources $\Omega=\{\omega_1,...,\omega_n\}$ and the mapping relations between $O$ and $\Omega$}
\Output{trust value $\tau(o_{i})$, $o_{i}\in O $; trustworthiness of source $t(\omega_j)$, $\omega_j\in \Omega $}
 Initialize the prior belief of all nodes $\psi_i(y_i)$; \\
 $\forall o_{i}, o_{j}\in O$: Calculating their similarity $S(o_i,o_{j})$;  \\
  $\forall y_{i}, y_{j}\in  Y$: $m_{i\rightarrow j}(y_i)=1$; //\textit{Initialize the message}\\
 \Repeat{the convergence criterion is satisfied}{
//\textit{Message propagation }\\
  \For { $j\leftarrow 1$ to $m+n$}{
 	\For{$i\leftarrow 1$ to $m+n$}{
 	$m_{i\rightarrow j}(y_j)\text{=}\sum_{y_i\in \{0,1\}} U(y_i,y_j)\psi_i(y_i)
        \prod_{y_k\in N(y_i) \cap  Y\setminus \{y_j\} }m_{k\rightarrow i}(y_i).$
 		  }
 		  }
}
\For {$i\leftarrow 1$ to $m+n$}{  
//\textit{Belief assignment}\\
 		 $P(y_i) =\psi_i(y_i) \prod_{y_j\in N(y_i) \cap  Y}m_{j\rightarrow i}(y_i). $ \\
 		  }
\Return  $\tau(o_{i}),\forall o_i\in O$; $t(\omega_j),\omega_j\in \Omega $
 \caption{ObResolution}
 \label{algo}
\end{algorithm}

\subsection{Practical Issues}
In this section, we discuss several important issues, including similarity functions and missing values, to ensure the practicality of our method. 

\textbf{Similarity functions.} The energy function between objects depends on the similarity function. We respect the characteristic of each data type and adopt different similarity functions to describe the similarity degrees. We discuss two similarity functions for numerical and categorical data, which are the two most common data types.

For numerical data, the most commonly used similarity function is defined as:

\begin{equation}\label{eq3} 
S(o_i,o_k) =1/1+d(o_i,o_k),
\end{equation}
\begin{equation}\label{eq4}
d(o_i,o_k) =
\left\{\begin{matrix}
 1 & if\ o_i=o_k=0,\\
{|o_i-o_k|}/max(|o_i|,|o_k|) & others. 
\end{matrix}\right.
\end{equation}

For string data, the Levenshtein distance \cite{32} is adopted to describe the similarities of objects. The similarity function is defined as follows:

\begin{equation}\label{eq5} 
S(o_i,o_k) =1-ld(o_i,o_k)/max(len(o_i),len(o_k)),
\end{equation}

\noindent where $ ld(o_i,o_k)$ denotes the Levenshtein distance between objects $o_i$ and $o_k$; $len(o_i) $ and $len(o_k) $ are the length of $o_i$ and $o_k$ respectively.

Apart from these functions, different similarity functions can be easily incorporated into our method to recognize the characteristics of various data types. We have obtained a few similarity functions in this study, which are selected based on data types. One of them is the Jaro-Winkler string similarity functions for names of people and strings that involve abbreviations \cite{jaro1989advances}.

\textbf{Missing Values}. Linked Data are built on the Open World Assumption, which states that what is not known to be true is simply unknown. Therefore, for the sake of simplicity in this study, we assume that all missing values are not known to be true.

\section{Evaluation}
\label{sect:eval}

\subsection{The Datasets}
Six Linked Data datasets were used in our experiments. The first three datasets \textit{persons}, \textit{locations}, \textit{organizations} are constructed based on the OAEI2011 New York Times dataset\footnote{http://data.nytimes.com/\#}, which is a well-known and carefully created dataset of Linked Data. In order to draw more robust conclusions, three other domains, including \textit{films}, \textit{books} and \textit{songs} are constructed through SPARQL queries over DBpedia. The construction process of datasets mainly involves the following two necessary steps.
\begin{enumerate}
\item\textbf{Identity Subjects}: We adopted a well-known tool, sameas.org\footnote{http://sameas.org/}, to identify subjects for the same real-world entities in the six dataset. Then, we crawled the data of every subject from BTC2014 \cite{btc-2014}, which is a comparatively complete LOD cloud tnat consists of 4 billion triples. \\
\item\textbf{Schema Mapping}: We adopted a method combining automatic matching and manual annotation to produce more accurate schema mapping results. First, features (\textit{Property Similarity}, \textit{Value Overlap Ratio},\textit{ Value Match Ratio} and \textit{Value Similarity Variance}) are selected based on the description provided in \cite{wang2015effective}. These selected features can achieve good performance in Linked Data. Subsequently, we chose the Random Forest and Support Vector Machine model, which achieved the best F1-Measure in \cite{wang2015effective} as classifiers for schema matching. Manual annotation is used to break the tie when an agreement was unreachable on a predicate between the two classifiers.
\end{enumerate} 

The statistics of the six datasets are shown in Table 3.  In this paper, an entity refers to something that has a real existence, such as a location.  A subject is a URIs that identifies the entity and different sources adopt various subjects to denote the same real-world entity (e.g., \textit{dbpedia:Statue of Liberty} and \textit{freebase:m.072p8}).  The ``\#ConflictingPredicates" refers to the number of subject/predicate pairs for which conflicting objects exist.
\vspace{-15pt}
\begin{table}[h]
	\caption{Statistics of the six datasets}
	\centering
	\begin{tabular}{lcccc}
		\hline
Datasets & \# Entities & \#Subjects & \#ConflictingPredicates& \#Triples\\
	\hline\hline		
 Persons& 4,978 & 21,340& 69,706& 141,937\\
 Locations& 1,910& 21,324& 38,200& 558,773\\
 Organizations& 2,000& 4,529& 14,000& 15,928\\
 Films& 2,000& 4,935& 8,000& 20,692\\ 
 Books&9,081& 15,644& 45,405& 71,532\\
 Songs&2,000& 2,872& 10,000& 7,170\\
 \hline
	\end{tabular}
\end{table}

One truth was selected from multiple conflicting objects for experimental verification. A strict process was established to ensure the quality of the annotation. This process mainly involved the following steps:
\begin{enumerate}
\item The annotators were provided annotated examples and annotation guidelines.
\item Every two annotators were asked to label the same predicate on the same entity independently.
\item The annotation results from two annotators were measured by using Cohen's kappa coefficient \cite{44}. The agreement coefficient of the six datasets was set to be at least 0.75. When an agreement could not be reached, a third annotator was asked to break the tie.
\end{enumerate}

The manually labeled results were regarded as the ground truth used in the evaluation.
\subsection{Comparative Methods and Metrics}
We compared our method with five well-known state-of-the-art truth discovery methods as competitors, which were modified, if necessary. 
\begin{itemize}
\item\textit{Majority Voting.} This method regards the object with the maximum number of occurrences as truth. Moreover, voting is a straightforward method.
\item\textit{Sums (Hubs and Authorities)} \cite{kleinberg1999authoritative}. This method regards the object supported by the maximum number of reliable sources as true. In this study, a source is recognized as a reliable source if its trustworthiness score exceeds 0.5.
\item\textit{TruthFinder} \cite{11}. This is a seminal work that is used to resolve conflicts based on source reliability estimation. It adopts Bayesian analysis to infer the trustworthiness of sources and the probabilities of a value being true.
\item\textit{ACCUCOPY} \cite{21}. This method is a popular truth discovery algorithm that obtains the highest precision among all methods in \cite{8}. ACCUCOPY considers the copying relationships between the sources, the accuracy of data sources, and the similarity between values.
\item\textit{F-Quality Assessment} \cite{4}. This method is a popular algorithm used to resolve conflicts in Linked Data. Three factors, namely the quality of the source, data conflicts, and confirmation of values from multiple sources, are leveraged for deciding which value should be the true value.
\end{itemize}

In the experiments, accuracy as a unified measure is adopted  and can be measured by computing the percentage of matched values between the output of each method and ground truths. The parameters of the baseline methods were set according to the authors' suggestions. We implemented all algorithms using Eclipse (Java) platform\footnote{https://www.eclipse.org/} by a single thread and conducted experiments on a windows sever computer with Intel Core E7-4820 CPU 2 GHz with 32 GB main memory, and Microsoft Windows 7 professional operating system. 

\subsection{Results}
\textbf{Accuracy Comparison.} Figure 2 shows the performance of different methods on the six datasets in terms of accuracy. As shown in this figure, our method consistently achieves the best accuracy among all the methods. Majority voting achieves the lowest accuracy (ranging from 0.3 to 0.45) on the six datasets among all the methods. Majority voting performs poorly in Linked Data for two reasons. First, approximately 50\% of predicates have no dominant object \cite{liu2016truthdiscover}. In this case, majority voting can only randomly select one object in order to break the tie. Second, majority voting assumes that all sources are equally reliable and does not distinguish them, which is not applicable to Linked Data as discussed in Introduction. Although source reliable estimation was taken into consideration in Sums, this method still achieves relatively low accuracy in all datasets because it only considers the correlation between sources and objects, but ignores the correlation between objects and the correlation between sources. TruthFinder, F-Quality Assessment and ACCUCOPY model all the clues by the iterative procedure, which easily leads to the problem of the rich getting richer over iterations. In this study, our proposed method utilizes the Source-Object network and successfully captures all the correlations from objects and Linked Data sources in a unified framework to infer the true objects, which explains why our method consistently achieves the best accuracy among all the comparative methods.
\vskip -15pt
\begin{figure}[!htb]
\centering
\includegraphics[width=0.8\textwidth]{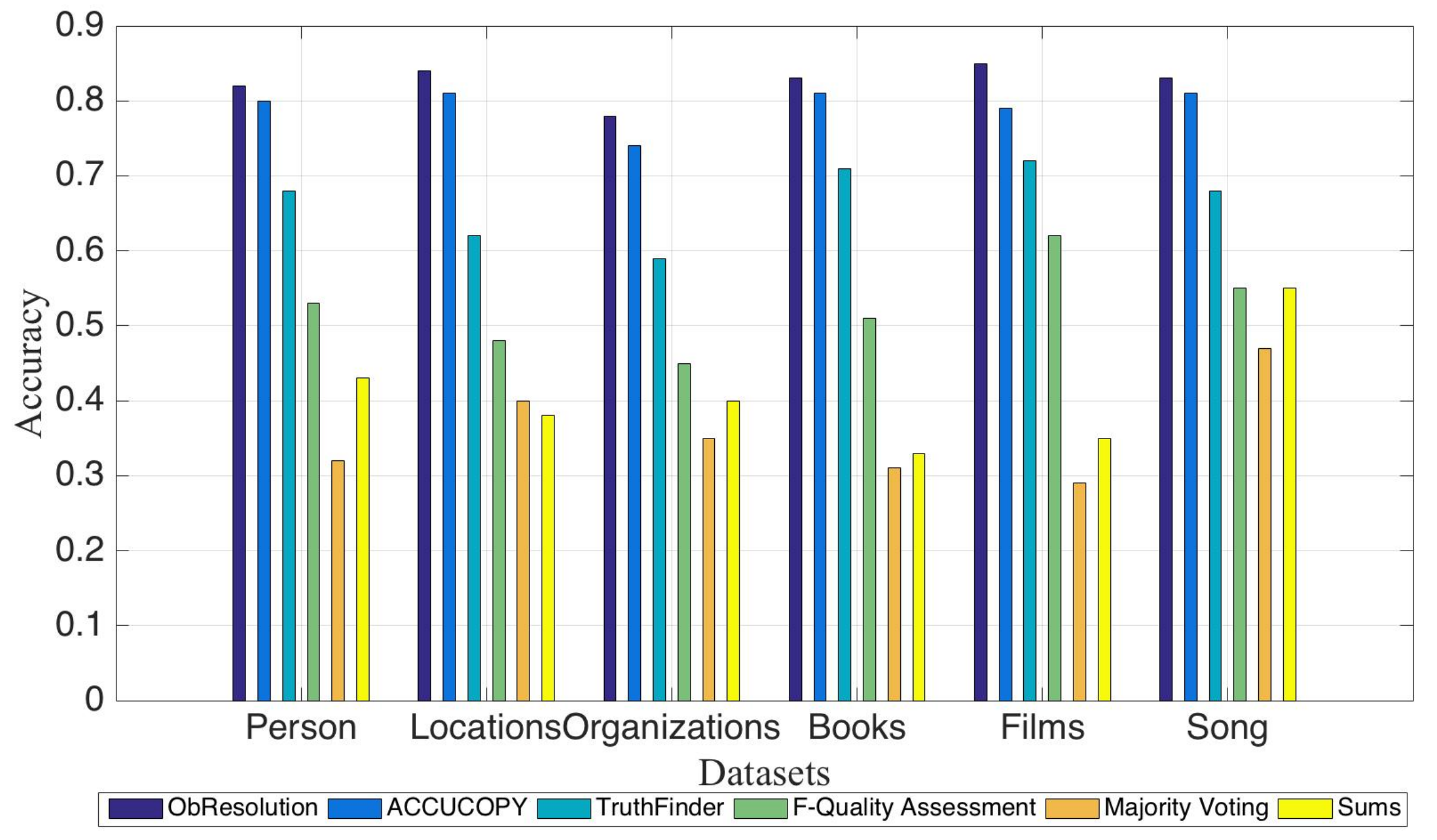}
\caption{Accuracy comparison in six datasets}
\end{figure} 
\vskip -15pt

\textbf{Sensitive Analysis. }We also studied the effect of the parameter $\beta, \delta$ on our methods. As discussed in Section 3, $\beta$ indicates the likelihood between reliable source and true object, whereas $\delta$ denotes the likelihood between unreliable source and false object. Figure 3 shows that the accuracy of ObResolution varies in different values of $\beta, \delta$ in the same dataset, and ObResolution achieves best accuracy on six datasets with different values of $\beta, \delta$ ($\beta=0.9, \delta=0.7$ for \textit{Persons}, for \textit{Books} $\beta=0.7, \delta=0.9$). Therefore, parameters $\beta, \delta$ are sensitive to different datasets because different Linked Data datasets have different quality \cite{7}. ObResolution uses different $\beta, \delta$ for different datasets to optimize the performance of our method.

\vskip -15pt
\begin{figure}[!htb]
\centering
\includegraphics[width=1\textwidth]{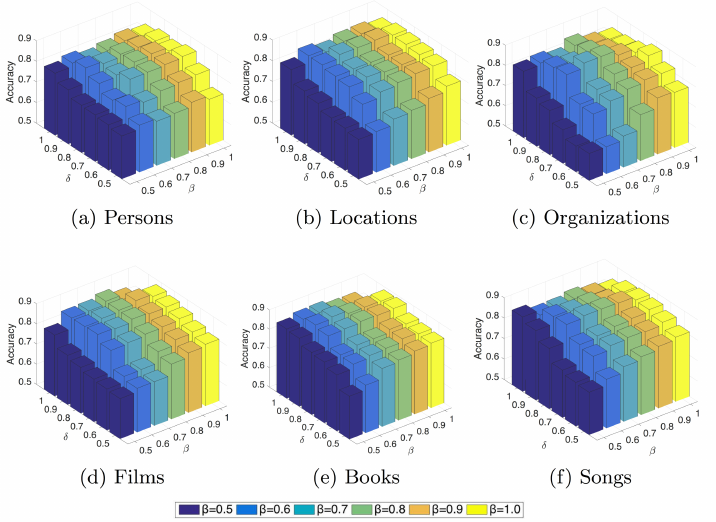}
\caption{Sensitive analysis in six Linked Data datasets}
\end{figure} 
\vskip -15pt
\section{Related Work}
\label{sect:work}

Resolving object conflicts is a key step for Linked Data integration and consumption. However, to the best of our knowledge, research on resolving object conflicts has not elicited enough attention in the Linked Data community. According to our survey, existing methods to resolve object conflicts in Linked Data can be grouped into three major categories of conflict handling strategies: conflict ignoring, conflict avoidance and conflict resolution.
\begin{itemize}
\item The conflict-ignoring strategy ignores the object conflicts and defers conflict resolution to users. For instance, Wang et al. \cite{wang2015effective} presented an effective framework to fuse knowledge cards from various search engines. In this framework, the fusion task involves card disambiguation and property alignment. For the value conflicts, this framework only adopted deduplication of the values and grouped these values into clusters.
 
\item The conflict-avoidance strategy acknowledges the existence of object conflicts, but does not resolve these conflicts. Alternatively, they apply a unique decision to all data, such as manual rules. For instance, Mendes et al. \cite{29} presented a Linked Data quality assessment framework called Sieve. In this framework, the strategy ``Trust Your Friends," which prefers the data from specific sources, was adopted to avoid conflicts. 
 
\item The conflict-resolution strategy focuses on how to solve a conflict regarding the characteristics of all data and metadata. For example, Michelfeit et al. \cite{4} presented an assessment model that leverages the quality of the source, data conflicts, and confirmation of values for determining which value should be the true value.
\end{itemize}

Previous work enlightens us on resolved object conflicts. In this paper, we propose a novel method that exploits the heterogeneous information network effect among sources and objects. Our approach is different in two aspects. First, we formalize the object conflict resolution problem through a heterogeneous information network, which successfully captures all the correlations from objects and Linked Data sources. Second, we adapt a message propagation-based method that exploits the network structure to infer the trust values of all objects. Our method has the following advantages: (1) it avoids the problem of the rich getting richer over iterations, and (2) it works in an unsupervised situation.

\section{Conclusion and Future Work}
\label{sect:concl}

Solving the problem of object conflicts is crucial to obtain insightful knowledge from a large number of Linked Data sources generated by numerous data contributors. In this paper, two objects are regarded as conflicts only when their similarity is less than the defined threshold. The two objects are still regarded as conflicts although they are expressed in terms of different measurement units. The main application scenario in this study are ranking the trust values of all objects and providing the most common ones for users rather than removing them directly. Therefore, this definition is reasonable in this sense. 

Existing studies on object conflicts either consider only the partial correlations from sources and objects, or model all clues by an iterative procedure. Differently, we proposed a novel method, ObResolution, to model all the clues from sources and objects in a unified framework using a heterogeneous information network called the Source-Object Network. In our method, the Source-Object Network was represented by pMRF because the trust values of all the nodes in this network are dependent on their neighbors and independent of all the other nodes. Thus, the problem of inferring the trust values of conflicting objects and trustworthiness of sources was defined as computing the joint distribution of variables. As such, we built a message propagation-based method that exploits the network structure to infer the trust values of all objects, and the object with the maximum trust score is regarded as the true object. We conducted experiments on six datasets collected from multiple platforms and demonstrated that ObResolution exhibits higher accuracy than several comparative methods.

A potential direction for future research is to address more complicated conflict resolution scenarios, such as the situation involving copying relations of different sources. Another future research direction is to investigate the case where a certain predicate of a real-world entity has several true values.\\

\noindent\textbf{Acknowledgments.} This work is funded by the National Key Research and Development Program of China (Grant No. 2016YFB1000903), the MOE Research Program for Online Education (Grant No. 2016YB166) and the National Science Foundation of China (Grant Nos. 61370019, 61672419, 61672418, 61532004, 61532015).


\end{document}